\documentclass[amsmath,11pt]{article}

\usepackage{amssymb,amssymb,epsfig,bm}
\usepackage{caption}
\usepackage{subcaption}
\usepackage{xcolor}
\date{\empty}

\begin{document}

\title{\bf General relativity and the bulk-flow puzzle}
\author{Christos G. Tsagas\\ {\small Section of Astrophysics, Astronomy and Mechanics, Department of Physics}\\ {\small Aristotle
University of Thessaloniki, Thessaloniki 54124, Greece}\\ {\small and}\\ {\small Clare Hall, University of Cambridge, Herschel Road, Cambridge CB3 9AL, UK}}

\maketitle

\begin{abstract}
Bulk peculiar flows are commonplace in the universe, with many surveys reporting their presence on scales spanning between few hundred and several hundred Mpc. However, the sizes and the speeds of some of these bulk flows are well in excess of those theoretically anticipated, which has made them a potentially serious problem for the $\Lambda$CDM model. Having said that, essentially all the available theoretical studies are Newtonian, or quasi-Newtonian, in nature and both bypass a key feature of peculiar motions, namely the gravitational contribution of the \textit{peculiar flux}. To begin with, recall that bulk flows are matter in motion and that moving matter means nonzero energy flux. In relativity energy fluxes gravitate, but the gravitational input of the peculiar flux has been largely bypassed. As we will show here, when the flux contribution to the gravitational field is accounted for, linear peculiar velocities grow considerably faster than in the Newtonian/quasi-Newtonian studies. Therefore, general relativity, could naturally relax the current $\Lambda$CDM limits to accommodate the reported fast bulk flows.
\end{abstract}

\section{Introduction}\label{sI}
%%%%%%%%%%%%%%%%%%%%%%%%%%%%%%%%
An open question, which is treated as a potential problem for the concordance cosmological model, are the observed bulk peculiar flows. Many surveys have repeatedly confirmed the presence of these large-scale motions, with typical sizes and speeds of few hundred Mpc and few hundred km/sec respectively~\cite{Aetal}. Although many of the reported bulk flows are within the $\Lambda$CDM requirements (e.g.~see~\cite{ND}), there is an increasing number of surveys reporting sizes and speeds in excess (or even well in excess) of the standard model (e.g.~see~\cite{HSLB}-\cite{CMSS}). Among them are the recent surveys of \cite{Wetal,WHD}, which used the \textit{CosmicFlows-4} data to report bulk flows considerably faster than anticipated. Interestingly, the surveys that agree with the $\Lambda$CDM report bulk flows on relatively small scales, roughly up to 100$\,h^{-1}$~Mpc, while those that disagree extend beyond the aforementioned threshold. Then, taking the latter reports at face value and assuming that the systematics are not responsible for the discrepancies, one wonders whether such fast and deep bulk peculiar motions could seriously undermine the $\Lambda$CDM paradigm.

Before resorting to drastic measures, however, it might help to take a step back and look closer at the theoretical models used to predict the features of the observed bulk flows. These are believed to have started as weak peculiar-velocity perturbations around recombination, when structure formation begun in earnest. It is the increasing inhomogeneity of the post-recombination universe that triggered, sustained and amplified the initial velocity perturbations to the bulk flows observed today. Nevertheless, the picture seems incomplete, since the theoretically predicted velocities are considerably slower than those reported in~\cite{HSLB}-\cite{WHD}. Having said that, the $\Lambda$CDM predictions are based entirely on Newtonian studies, which by default bypass the gravitational input from a key feature of peculiar motions, namely from their energy flux.

Peculiar flows are nothing else but matter in motion and moving matter carries energy flux. In relativity, as opposed to Newtonian physics, energy fluxes contribute to the energy-momentum tensor and therefore to the local gravitational field  (e.g.~see~\cite{TCM,EMM}). Without accounting for the gravitational input of the peculiar flux, the Newtonian studies have led to the relatively moderate growth rate of $v\propto t^{1/3}$ for the linear peculiar velocity field ($v$) after equipartition (e.g.~see~\cite{Pe}). The literature also contains a few quasi-Newtonian linear treatments that recover the Newtonian growth-rate (see~\S~\ref{ssLQ-NA} below). However, despite their relativistic initial appearance, these studies also reduce to Newtonian, because they also bypass the (purely general relativistic) contribution of the peculiar flux to the gravitational field. Although this is unavoidable for the purely Newtonian studies, in their quasi-Newtonian analogues the reason is the severe restrictions imposed upon the perturbed spacetime. These constraints may simplify the calculations, but their repercussions seriously compromise the relativistic nature of the study.

It should be noted that the problematic nature of the highly restrictive quasi-Newtonian setup has been known and noted, although not widely, at least since~\cite{EMM}. The interested reader is referred to \S~6.8.2 in~\cite{EMM} for a discussion and for ``warning''  comments, as well as to \S~\ref{ssLQ-NA} here. However, the extent of the problem has not been realised, which explains why linear quasi-Newtonian studies of peculiar-velocity fields are still misleadingly termed relativistic. The reason is probably because, so far, there has been no direct comparison between the quasi-Newtonian and the proper relativistic study. It is one of the aims of this work to make the comparison and in so doing demonstrate the extent of the problem.\footnote{There are more examples of studies involving peculiar motions, which start relativistically and reduce to Newtonian when the flux-input of the moving matter to the gravitational field is bypassed (see \S~\ref{sFRPF} below).}

In a proper relativistic linear analysis of cosmological peculiar velocities there are no quasi-Newtonian restrictions, for the simple reason that the resulting simplifications are not necessary. The real difference comes from the role of the peculiar flux and whether its gravitational input to Einstein's equations is accounted for, or not. The flux reflects the fact that, when peculiar motions are present, the cosmic fluid cannot be treated as perfect, even at the linear level. The ``imperfection'' appears as a nonzero \textit{peculiar flux} due to the moving matter (e.g.~see \S~\ref{sPFP4A} here and \S~5.2.1 in~\cite{EMM}). Less well known is that the 4-acceleration is also nonzero even in the absence of pressure. Put another way, the peculiar flux is always linked to a \textit{peculiar 4-acceleration}. In addition, the peculiar-flux contribution to the relativistic gravitational field feeds, through Einstein's equations, into the conservations laws and eventually appears in the formulae monitoring the evolution of linear peculiar velocities. The result is the considerably stronger linear growth-rate of $v\propto t$ for the peculiar-velocity perturbations. Moreover, the theory of differential equations guarantees that the aforementioned growth-rate is the minimum possible~\cite{TT}-\cite{MiT}. The overall message is that general relativity leads to substantially stronger growth rates for the linear peculiar-velocity field and, in so doing, it can provide a theoretical answer to the question raised by the fast and deep bulk flows reported by many surveys. Moreover, this could be achieved within the framework of the $\Lambda$CDM paradigm.

\section{Peculiar flux and peculiar 4-acceleration}\label{sPFP4A}
%%%%%%%%%%%%%%%%%%%%%%%%%%%%%%%%%%%%%%%%%%%%%%%%%%%%%%%%%%%%%%%%%
Cosmological peculiar motions require a universal reference system, relative to which one can define and measure them. Typically, this is the rest-frame of the Cosmic Microwave Background (CMB), which is defined as the only coordinate system where the radiation dipole vanishes~\cite{EMM}.

\subsection{The peculiar flux}\label{ssPF}
%%%%%%%%%%%%%%%%%%%%%%%%%%%%%%%%%%%%%%%%%%
Relativistic studies of peculiar motions require ``tilted'' spacetimes, with two groups of observers in relative motion. Here, we consider a tilted, perturbed Friedmann-Robertson-Walker (FRW) universe with two 4-velocity fields $u_a$ and $\tilde{u}_a$. The former is the reference (CMB) frame of the universe and the latter is that of the moving matter. For non-relativistic peculiar motions, the two 4-velocities are related by $\tilde{u}_a=u_a+ v_a$, where $v_a$ represents the relative velocity of the matter (with $u_av^a=0$ and $v^2\ll1$). Note that both frames ``live'' in the perturbed universe (recall that the CMB is nearly but not fully isotropic) and that none of the quasi-Newtonian restrictions are imposed on either them~\cite{TT}-\cite{MiT}.\footnote{In the FRW background the $u_a$ and the $\tilde{u}_a$ fields coincide by default, which makes the peculiar velocity vector ($v_a$) a gauge-invariant linear perturbation~\cite{SW}.}

Relative motions affect the type of matter experienced by the relatively moving observers. In particular, when the background spacetime is described by one of the Friedmann models, the energy density ($\rho$), the (isotropic) pressure ($p$), the energy flux ($q_a$) and the viscosity ($\pi_{ab}$) of the matter, as measured in the two frames, are related by
\begin{equation}
\tilde{\rho}= \rho\,, \hspace{10mm} \tilde{p}= p\,, \hspace{10mm} \tilde{q}_a= q_a- (\rho+p)v_a \hspace{5mm} {\rm and} \hspace{5mm} \tilde{\pi}_{ab}= \pi_{ab}\,,  \label{lrels1}
\end{equation}
to first approximation.\footnote{Hereafter tilded variables will always be  measured in the coordinate system of the matter, while their non-tilded counterparts will be evaluated in the reference CMB frame.} Accordingly, if the pressure (both isotropic and viscous) is zero in one frame, it vanishes in any other relatively moving coordinate system (at the linear level). This is also the case for the density, but not for the energy flux. Indeed, setting $q_a=0$ in the CMB frame leads to $\tilde{q}_a=-(\rho+p)v_a$ in that of the matter, while $\tilde{q}_a=0$ means that $q_a=(\rho+p)v_a$. Clearly, setting both flux-vectors to zero in Eq.~(\ref{lrels1}c) leaves no peculiar-velocity field to study. The only exception is when the matter has a (de Sitter) inflationary equation of state with $p=-\rho$ (see~\cite{MaT} and also \S~\ref{ssTOs} below). Therefore, following (\ref{lrels1}c), the relative motion between the two frames ensures that there is always a nonzero linear flux vector in the system. Put another way, the cosmic medium can no longer be treated as perfect. The ``imperfection'' appears in the form of a nonzero \textit{peculiar flux} vector solely triggered by relative-motion effects (e.g.~see \S~5.2.1 in~\cite{EMM}).

In section \S~\ref{sRQ-NSs}, we will compare the relativistic study of linear peculiar velocities to the quasi-Newtonian treatments of~\cite{M}. We will therefore adopt the conventions of these papers to facilitate the comparison. More specifically, we will assume an Einstein-de Sitter background and set the flux and the 4-acceleration to zero in the matter frame (i.e.~set $\tilde{q}_a=0$ and $\tilde{A}_a=0$ respectively). So, there is no peculiar flux in the coordinate system of the pressureless matter, which moves along timelike geodesics. Then, in the absence of pressure, relation (\ref{lrels1}c) leads to
\begin{equation}
q_a= \rho v_a\,,  \label{lpf}
\end{equation}
which is the flux triggered by the motion of the matter relative to the rest-frame of the CMB photons.

\subsection{The peculiar 4-acceleration}\label{ssP4-A}
%%%%%%%%%%%%%%%%%%%%%%%%%%%%%%%%%%%%%%%%%%%%%%%%%%%%%%
In relativity, as opposed to Newtonian gravity, energy fluxes also contribute to the energy-momentum tensor. Therefore, in a sense, one could say that peculiar motions ``gravitate''~\cite{TT}. Through Einstein's equations, the gravitational input of the peculiar flux feeds into the relativistic conservations laws, which acquire flux-related terms. More specifically, the linearised energy and momentum conservation laws respectively read
\begin{equation}
\dot{\rho}= -\Theta\rho- {\rm D}^aq_a \hspace{15mm} {\rm and} \hspace{15mm} \rho A_a= -\dot{q}_a- 4Hq_a\,,  \label{lcls}
\end{equation}
where $\Theta={\rm D}^au_a>0$ is the expansion scalar, $H$ is the background Hubble parameter (with $\Theta=3H$ there) and ${\rm D}_a$ is the 3-D (covariant) derivative operator~\cite{TCM,EMM}. Hence, despite the absence of pressure, the 4-acceleration ($A_a$) is not zero.\footnote{Following (\ref{lcls}b), zero 4-acceleration is compatible with nonzero flux only when $\dot{q}_a=-4Hq_a$, namely in one out of theoretically infinite possibilities (a set of measure zero probabilistically). Moreover, this one and only case leads to a linear $v_a$-field that decays as $v_a\propto a^{-1}\propto t^{-2/3}$, which is cosmologically unacceptable. Indeed, since peculiar velocities start weak at around recombination, there should be no bulk flows to observe today, if they were to decay with time. Note that, even in the Newtonian studies, linear peculiar velocities grow as $v_a\propto t^{1/3}$.} This \textit{peculiar 4-acceleration} also reflects the fact that the cosmic medium is no longer perfect when peculiar motions are present.

\section{Relativistic vs quasi-Newtonian studies}\label{sRQ-NSs}
%%%%%%%%%%%%%%%%%%%%%%%%%%%%%%%%%%%%%%%%%%%%%%%%%%%%%%%%%%%%%%%%
Despite its initially relativistic profile, the quasi-Newtonian approach imposes strict constraints upon the host spacetime that compromise its relativistic nature. Although the problem has been known (e.g.~see \S~6.8.2 in~\cite{EMM}), so far there has been no direct comparison with the fully relativistic studies to reveal the extent of it. Next, we will provide such a comparison by looking at the linear evolution of large-scale peculiar velocities.

\subsection{Linear quasi-Newtonian approach}\label{ssLQ-NA}
%%%%%%%%%%%%%%%%%%%%%%%%%%%%%%%%%%%%%%%%%%%%%%%%%%%%%%%%%%%
Quasi-Newtonian studies adopt a (reference) frame with zero linear vorticity and shear. The restrictions do not stop there, however, but lead to further compromising constraints. These deprive the perturbed spacetime from key relativistic features, like gravitational waves for example, eventually leading to Newtonian-like equations and results (see \S~6.8.2 in~\cite{EMM} for ``warning'' comments). When studying peculiar motions, the starting point is the linear evolution law
\begin{equation}
\dot{v}_a+ Hv_a= -A_a\,,  \label{ldotv}
\end{equation}
The apparent ``advantage'' of the quasi-Newtonian approach is that, without vorticity, one can appeal to a scalar potential ($\varphi$) and write the 4-acceleration as the gradient of an effective potential, by means of the expression~\cite{M}
\begin{equation}
A_a= {\rm D}_a\varphi\,,  \label{qNlA}
\end{equation}
which is identical to its purely Newtonian counterpart for all practical purposes. Note that $\varphi$ is an ad hoc potential, the time-evolution of which follows from the ansatz $\dot{\varphi}=-\Theta/3$. Moreover, the latter is not uniquely determined and requires setting the shear to zero as well. The worst side effect of (\ref{qNlA}), however, is that it does not account for the gravitational input of the peculiar flux and this omission severely compromises the study of peculiar velocities. It should therefore come to no surprise that both the Newtonian and the quasi-Newtonian treatments lead to the same differential equation for the linear evolution of the $v_a$-field, namely to~\cite{M}
\begin{equation}
\ddot{v}_a= -3H\dot{v}_a+ H^2v_a= -{2\over t}\,\dot{v}_a+ {4\over9t^2}\,v_a\,,  \label{lq-Nddotv}
\end{equation}
with the second equality reflecting the fact that $a\propto t^{2/3}$ and $H=2/3t$ after equipartition.\footnote{Equation (\ref{lq-Nddotv}) follows after employing the linear commutation law $({\rm D}_a\varphi)^{\cdot}={\rm D}_a\dot{\varphi}-H{\rm D}_a\varphi+\dot{\varphi}A_a$, imposing the ansatz $\dot{\varphi}=-\Theta$ and then using the zero shear assumption to obtain the constraint ${\rm D}_a\Theta=9H^2v_a/2$~\cite{M}.} As expected, the above accepts the power-law solution~\cite{M}
\begin{equation}
v= \mathcal{C}_1t^{1/3}+ \mathcal{C}_2t^{-4/3}= \mathcal{C}_3a^{1/2}+ \mathcal{C}_4a^{-2}\,,  \label{lqNv}
\end{equation}
which simply reproduces the purely Newtonian growth-rate reported in~\cite{Pe}. In retrospect, the agreement between the two approaches was unavoidable, since they have both bypassed (albeit for different reasons) the contribution of the peculiar flux to the gravitational field. As we will show in \S~\ref{ssLRA} next, it is the gravitational input of the peculiar-flux that separates the relativistic studies from the rest.\footnote{Both the zero shear and vorticity constraints of the quasi-Newtonian treatment, as well as the subsequent introduction of the scalar potential ($\varphi$), are not necessary. One can study linear peculiar velocities in full general relativity without any of the previous restrictions. Thus, from the relativistic viewpoint, the quasi-Newtonian study of peculiar-velocity fields is an interesting mathematical exercise but without real physical insight.}

\subsection{Linear relativistic approach}\label{ssLRA}
%%%%%%%%%%%%%%%%%%%%%%%%%%%%%%%%%%%%%%%%%%%%%%%%%%%%%%
There are more than one ways of showing that a relativistic study that accounts for the effects of the peculiar flux, leads to faster linear growth-rates for the peculiar-velocity field (e.g.~see~\cite{TT}-\cite{MiT}). In what follows, we will adopt an alternative and more direct approach that provides a clearer insight in our opinion.

Starting from the conservation law of the momentum density (\ref{lcls}b) and keeping in mind that $q_a=\rho v_a$ (see Eq.~(\ref{lpf}) in \S~\ref{ssPF}), it is straightforward to show that
\begin{equation}
A_a= -\dot{v}_a- Hv_a\,,  \label{rlAa}
\end{equation}
which agrees with its quasi-Newtonian counterpart (compare to Eq.~(\ref{ldotv}) in \S~\ref{ssLQ-NA} previously). However, the agreement between the quasi-Newtonian and the relativistic analysis of peculiar velocities stops at this point. The two approaches begin to diverge, because the relativistic 4-acceleration is no longer given by the quasi-Newtonian ansatz (\ref{qNlA}). Instead, taking the spatial gradient of the energy conservation law (\ref{lcls}a) and linearising, one arrives at the following expression for the 4-acceleration ($A_a$) in the presence of peculiar motions~\cite{TT}-\cite{MiT}
\begin{equation}
\dot{\Delta}_a= -\mathcal{Z}_a- 3aHA_a- a{\rm D}_a\vartheta\,.  \label{ldotDel}
\end{equation}
The spatial gradients $\Delta_a=(a/\rho){\rm D}_a\rho$ and $\mathcal{Z}_a=a{\rm D}_a\Theta$ describe inhomogeneities in the matter density and the universal expansion respectively (e.g.~see~\cite{TCM,EMM}). Also, $\vartheta={\rm D}^av_a$ is the 3-divergence of the peculiar-velocity field, reflecting the input of the latter's flux to the continuity equation (\ref{lcls}a). The above ensures that, in addition to the peculiar flux, the 4-acceleration also brings into play the density and the expansion gradients. The evolution of $\mathcal{Z}_a$ follows from the linearised spatial gradient of the Raychaudhuri equation, namely from
\begin{equation}
\dot{\mathcal{Z}}_a= -2H\mathcal{Z}_a- {1\over2}\,\rho\Delta_a- {9\over2}\,aH^2A_a+ a{\rm D}_a{\rm D}^bA_b\,.  \label{ldotcZ}
\end{equation}
Relations (\ref{ldotDel}) and (\ref{ldotcZ}) are also the linear propagation equations for the density gradients and the expansion gradients respectively in the presence of peculiar motions.\footnote{Alternatively, one can obtain (\ref{ldotDel}) and (\ref{ldotcZ}) by linearising the nonlinear formulae (2.3.1) and (2.3.2) of~\cite{TCM}, or Eqs.~(10.101) and (10.102) of~\cite{EMM}, while taking into account the linear relations (\ref{lpf}) and (\ref{lcls}b).} Therefore, the relativistic approach directly relates the peculiar 4-acceleration to the evolution of these two fields as well. This is the result of accounting for the gravitational contribution of the peculiar flux, namely for a purely general-relativistic input that is unaccounted for in both the Newtonian and the quasi-Newtonian approaches (though for different reasons).

Taking the time derivative of (\ref{ldotDel}), substituting (\ref{ldotcZ}) into the resulting expression and using relation (\ref{rlAa}), we arrive at the following linear propagation equation for the peculiar-velocity field
\begin{equation}
\ddot{v}_a+ H\dot{v}_a- {3\over2}\,H^2v_a= {1\over3aH}\left(\ddot{\Delta}_a+ 2H\dot{\Delta}_a- {3\over2}\,H^2\Delta_a\right)\,.  \label{lddotv1}
\end{equation}
In deriving the above, we have also used the background relation $\dot{H}=-3H^2/2$, together with the linear commutation laws $({\rm D}^av_a)^{\cdot}={\rm D}^a\dot{v}_a- H{\rm D}^av_a$ and $({\rm D}_a\vartheta)^{\cdot}={\rm D}_a\dot{\vartheta}-H{\rm D}_a\vartheta$. Expression (\ref{lddotv1}) is a non-homogeneous differential formula, relating the linear evolution of peculiar-velocity perturbations to those in the density distribution of the pressure-free matter, in a tilted almost-FRW universe with zero spatial curvature.\footnote{It should be noted that the terms non-homogeneous/homogeneous refer only to the nature of the differential equations and not to the homogeneity of the space, which is both inhomogeneous and anisotropic at the linear perturbative level.} Note that the pressureless nature of the matter implies that (\ref{lddotv1}) applies to baryonic ``dust'' after recombination, as well as to low-energy Cold Dark Matter (CDM) species after equipartition. Expression (\ref{lddotv1}) is the (fully relativistic) differential equation governing linear peculiar velocities in a tilted Einstein-de Sitter universe. Next, we will show that the minimum growth-rate of the peculiar-velocity field predicted by the relativistic analysis, is considerably stronger than the Newtonian/quasi-Newtonian rate.

Our starting point is a well known theorem of linear differential equations, according to which the full solution of a non-homogeneous linear differential equation forms from the general solution of its homogeneous component and from a partial solution of the full equation. Let us apply the above theorem to Eq.~(\ref{lddotv1}) and isolate its homogeneous part, which reads
\begin{equation}
\ddot{v}_a+ {2\over3t}\,\dot{v}_a- {2\over3t^2}\,v_a= 0\,,  \label{hlddotv1}
\end{equation}
given that $H=2/3t$ after matter-radiation equality. The above solves analytically to give
\begin{equation}
v= {1\over5}\left(2v_0+3\dot{v}_0t_0\right)\left({t\over t_0}\right)+ {3\over5}\left(v_0-\dot{v}_0t_0\right) \left({t_0\over t}\right)^{2/3}\,,  \label{lv1}
\end{equation}
with $v_0$ and $\dot{v}_0$ determined by the initial conditions (at $t=t_0$). Therefore, solving the homogenous part of (\ref{lddotv1}) led to the growth-rate of $v\propto t$ for the linear peculiar-velocity field, which is considerably stronger than the $v\propto t^{1/3}$ growth of the Newtonian and quasi-Newtonian treatments (see \S~\ref{ssLQ-NA} earlier). Moreover, it is important to realise that $v\propto t$ is also the minimum linear growth-rate. Indeed, in line with the aforementioned theorem on differential equations, the full solution of the non-homogeneous Eq.~(\ref{lddotv1}) forms from the general solution of its homogeneous part (i.e.~from (\ref{lv1}) above) and from a partial solution of the original non-homogeneous differential formula. Therefore, solving (\ref{lddotv1}) in full will make physical difference only if the partial solution grows faster than the fastest growing mode of its homogeneous counterpart.\footnote{Although rather unlikely, it is conceivable that the partial solution of the non-homogeneous differential equation (\ref{lddotv1}) could cancel out the growing mode of the homogeneous solution (\ref{lv1}). However, this will probably require a unique fine-tuning of the initial conditions.} Put another way, mathematically speaking, the theory of differential equations guarantees that $v\propto t$ is the minimum growth-rate for the linear peculiar-velocity field.

The physics goes along with the mathematics and perhaps the most straightforward and intuitive way to demonstrate this is by involving the 4-acceleration ($A_a$), namely the driving force of the peculiar-velocity field (see Eq.~(\ref{rlAa}) earlier). This means that the stronger the 4-acceleration, the faster the peculiar velocity. Following Eq.~(\ref{ldotDel}) and the comments immediately after, in addition to the effects of the peculiar flux, $A_a$ also brings into play those of the density and the expansion gradients. Therefore, the overall impact of the driving force depends on which and how many of these agents are included in the final solution, thus allowing for studies that can complement each other.

Without the non-homogeneous right-hand side of differential equation (\ref{lddotv1}), the $v\propto t$ growth-rate reported in this work (see solution (\ref{lv1}) above) accounts only for the flux effects and it is driven by a constant 4-acceleration (consult Eq.~(\ref{rlAa}) for a quick check). On the other hand, the faster $v\propto t^{4/3}$ rate of the earlier relativistic studies also accounts for some of the gradient effects and for this reason it is driven by a growing 4-acceleration (with $A_a\propto t^{1/3}$ at a minimum - see~\cite{TT} and also Eq.~(\ref{rlAa}) here). It is conceivable that this rate could increase further, if all the gradient effects were to be accounted for. In contrast, none of the aforementioned agents contributes to the quasi-Newtonian 4-acceleration, which decays as $A_a\propto t^{-2/3}$ (see~\cite{M}, or consult Eq.~(\ref{rlAa}) here). This drastic change in the evolution of the 4-acceleration explains why the quasi-Newtonian treatment leads to $v\propto t^{1/3}$ and thus to slower peculiar velocities. Clearly, the latter applies to the purely Newtonian treatment as well.

In summary, even in the ``minimalist'' scenario, where the right-hand side of (\ref{lddotv1}) is bypassed and only the flux effects are accounted for, the relativistic analysis still leads to faster peculiar velocities than the Newtonian/quasi-Newtonian studies. This happens because the gravitational input of the peculiar flux enhances the linear growth of the peculiar-velocity field. As a result, relativity supports faster and deeper residual bulk flows, beyond the typical expectations of the $\Lambda$CDM model and perhaps like those reported in~\cite{CMSS} and~\cite{WHD} for example. Demonstrating this principle in a mathematically simple way is the main aim of this work. We chose to do so by using the long-known expressions (9) and (10), which are then easily combined to give differential equation (11).

In the recent literature there are also simulations of structure formation that use relativistic numerical techniques (e.g.~\cite{Adetal}). Some works also employ simulations to study the large-scale peculiar kinematics. In~\cite{Jetal} for example, the authors focused on the generation of vorticity. To the best of our knowledge, the simulations largely agree with the Newtonian picture. However, before comparing our analytical approach to the simulations, one should keep in mind that: (i) the majority of the simulations (including the last two) operate within the $\Lambda$CDM model, where, the accelerated expansion typically suppresses the growth of essentially all kinds of distortions, including the peculiar velocities (see \S~\ref{ssTOs} next); (ii) even during the preceding Einstein-de Sitter phase, the rotational component of the peculiar-velocity field (investigated in simulations such as those of~\cite{Jetal}) evolves considerably slower than the velocity field itself and the same also applies to the rest of the velocity gradients (i.e.~to the divergence and the shear -- see~\cite{TT}). It is very likely that the aforementioned structural differences, both of which inhibit (or even suppress) the impact of the simulated peculiar-velocity field, are the reasons for the apparent disagreement between the analytical and the numerical approaches. Put another way, as yet, there seems to be no clear common ground where one could directly compare the analytical work to the simulations and vice versa. On the positive side, the relativistic simulations are under development and there should still be room for further testing and improvement. Especially in view of the increasing number of surveys reporting bulk peculiar flows in excess of the Newtonian/$\Lambda$CDM predictions. Then, the analytical results presented here and their theoretical support to the aforementioned fast and deep bulk flows should provide the motivation for adapting/extending the existing numerical codes to address the ongoing bulk-flow puzzle, while offering a useful testing ground for them at the same time.

\subsection{Theory vs observations}\label{ssTOs}
%%%%%%%%%%%%%%%%%%%%%%%%%%%%%%%%%%%%%%%%%%%%%%%%
Within the framework of the $\Lambda$CDM paradigm, solution (\ref{lv1}) holds between decoupling and the onset of the accelerated phase, when applied to conventional baryonic dust. Alternatively, when dealing with peculiar-velocity perturbations in the distribution of the low-energy CMD species, solution (\ref{lv1}) holds between equipartition and the accelerated phase. During both of these periods, when the universe is believed to be close to the Einstein-de Sitter model, our results show faster linear growth for the peculiar-velocity field. Given that the growth rate of $v\propto t$ is also the minimum possible, it is fair ro argue that the relativistic analysis provides theoretical support to several surveys claiming peculiar velocities faster than anticipated~\cite{HSLB}-\cite{WHD}. Overall, general relativity can provide a simple and physically motivated answer to the bulk-flow puzzle.

\begin{figure}[!tbp]\vspace{-7.5mm}
  \begin{subfigure}[b]{0.475\textwidth}
    \includegraphics[width=\textwidth]{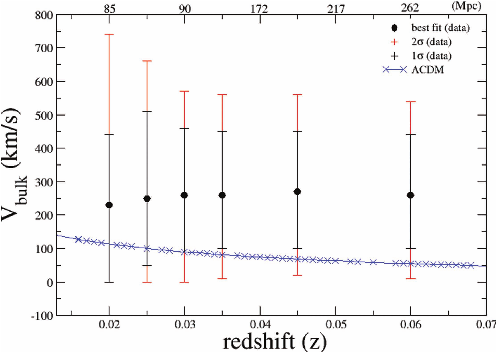}
    \caption{from Colin, et al~\cite{CMSS}}
    \label{fig:f1}
  \end{subfigure}
  \hfill
  \begin{subfigure}[b]{0.475\textwidth}
    \includegraphics[width=\textwidth]{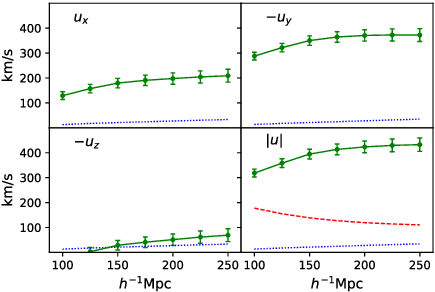}
    \caption{from Watkins, et al~\cite{Wetal}}
    \label{fig:f2}
  \end{subfigure}
  \caption{(a) The redshift profile of the bulk flow from the likelihood analysis of~\cite{CMSS}. The peculiar velocity (black dots) systematically exceeds the blue line of the $\Lambda$CDM expectations, but its value also shows signs of decrease at lower redshifts. Note that Fig.~\ref{fig:f1} shows that 1-D expectation, so it needs to be multiplied by $\sqrt{3}$ to give the 3-D one (see also Fig.~8 in~\cite{CMSS}). (b) The scale dependence of the bulk-flow velocity along the three coordinate axes and of its mean value (green lines), estimated from the \textit{CosmicFlows4} catalogue~\cite{Wetal}. The red dashed line is the theoretical expectation for the mean bulk velocity according to the standard cosmological model. Note the profound disagreement between the predicted and the measured bulk-flow velocities, as well as the decline in the magnitude of the latter at lower redshifts (see also Fig.~7 in~\cite{Wetal}).}  \label{fig:CetalWetal}
\end{figure}

At this point, we would like to draw the reader's attention to an additional issue of interest. Remaining within the $\Lambda$CDM scenario, the Einstein-de Sitter period of the universe is followed by a recent phase of accelerated expansion. Once cosmic acceleration starts, the growth of the $v$-field is expected to slow down (if not to decrease). For instance, when the accelerated expansion is driven by a scalar field with an effective $p=-\rho$ equation of state, linear peculiar-velocity perturbations were found to decay as $v\propto a^{-1}$ (see~\cite{MaT} for a discussion). In general, the strength of the effect depends on the rate of the acceleration, which in turn depends on the driving agent. The latter could have the form of a cosmological constant, or of some dynamically evolving dark energy, in which case one also needs to know the effective equation of state. In any case, the expansion starts to accelerate late into the Einstein-de Sitter epoch. Prior to that, the growing mode seen in solution (\ref{lv1}) will quickly dictate the evolution of the peculiar-velocity field, unless the initial conditions are fine-tuned against it.\footnote{In an exactly analogous way the growing mode of linear density perturbations quickly dominates and dictates their evolution.} Then, some (at least) of the earlier enhancement will survive through the accelerated phase to the present. On these grounds, one should expect to measure bulk velocities faster than anticipated, but also to see them decline at relatively low redshifts. Qualitatively speaking, such a peculiar-velocity profile closely resembles those reported in~\cite{CMSS} and in~\cite{Wetal} (see Figs.~\ref{fig:f1} and \ref{fig:f2} here). Then, the peculiar-velocity decline at low redshifts, as seen in Fig.~\ref{fig:CetalWetal}, may simply reflect the universe's late-time accelerated expansion. If so, the discrepancy between the reported fast and deep bulk flows and the $\Lambda$CDM predictions may not reflect a generic problem of the current cosmological model, but it may instead indicate the use of the inappropriate gravitational theory when studying large-scale peculiar motions.

\section{The fundamental role of the peculiar flux}\label{sFRPF}
%%%%%%%%%%%%%%%%%%%%%%%%%%%%%%%%%%%%%%%%%%%%%%%%%%%%%%%%%%%%%%%%
Without accounting for the gravitational input of the peculiar flux,  the quasi-Newtonian equations simply reproduced the purely Newtonian ($v\propto t^{1/3}$) growth-rate for the linear-peculiar velocity field (see solution (\ref{lqNv}) in \S~\ref{ssLQ-NA}). This was the result of imposing strict constrains on the perturbed spacetime, which compromised its relativistic nature. In contrast, without imposing any constraints and by accounting for the gravitational contribution of the peculiar flux, the relativistic analysis led to the considerably stronger growth ($v\propto t$) for linear peculiar velocities. This makes the peculiar flux the key physical agent that separates the Newtonian/quasi-Newtonian from the relativistic studies of peculiar motions in cosmology.

In the literature, there are additional examples of studies involving peculiar motions that start relativistically but end up Newtonian, when the gravitational effect of the peculiar flux is bypassed for one reason or another. The relativistic approach to the Zeldovich approximation of~\cite{ET}, in particular, reproduced the Newtonian ``pancake'' attractor, once the quasi-Newtonian frame was adopted and the 4-acceleration was replaced by the gradient of a potential, identical to that of~\cite{M}.  Another example is the relativistic treatment of the Meszaros ``stagnation'' effect, where linear perturbations in the density of the dust component ``freeze'' during the radiation era~\cite{Me}. This time no quasi-Newtonian frame was introduced, the gravitational input of the peculiar flux was accounted for and all the flux-related terms (as seen in Eqs.~(\ref{lcls}), (\ref{ldotDel}) and (\ref{ldotcZ}) here) were initially included into the linear equations. In the process, however, the analysis was switched to the Landau-Lifshitz (or energy) frame, where the flux vanishes by default (see \S~3.3.3 and \S~3.3.4 in~\cite{TCM}, or \S~10.4.3 in~\cite{EMM}). Without the flux input, the ``relativistic'' result was identical to the Newtonian solution of~\cite{Me}. It would be interesting to see what happens when the peculiar flux is properly accounted for in both of the aforementioned cases.

Our analysis, as well as the aforementioned characteristic examples, clearly demonstrate that studies that may have started as relativistic reduce to Newtonian for all practical purposes, when the gravitational input of the peculiar flux is (for one reason or another) unaccounted for.

\section{Discussion}\label{sD}
%%%%%%%%%%%%%%%%%%%%%%%%%%%%%%\\
In general relativity, moving matter has an additional input to the gravitational field, since its flux also contributes to the energy-momentum tensor. It is then only natural to argue that any study of peculiar motions, which claims to be relativistic, must account for the above effect. Otherwise, the study is in danger of severely compromising its relativistic nature and even reducing to Newtonian. All this is simple common sense and it should go without saying.

An example of matter in motion, on cosmological scales, are the observed bulk peculiar flows. These are believed to have started as weak velocity perturbations at recombination, which were subsequently amplified by structure formation. There are problems, however, because several recent surveys have reported bulk velocities well in excess of those expected. Having said that, the available theoretical studies are still few and sparse and they are almost all Newtonian. There are also few  quasi-Newtonian works that have the external appearance of a proper relativistic analysis. Nevertheless, by its own nature, the quasi-Newtonian framework is so restrictive that it eventually leads to Newtonian-like equations and results (e.g.~see \S~1.4.2 here and also \S~6.8.2 in~\cite{EMM}). As a result, both studies arrive at the same mediocre ($v\propto t^{1/3}$) growth-rate for the linear peculiar-velocity field (e.g.~see~\cite{Pe,M} and \S~\ref{ssLQ-NA} here), which is too weak to explain the reported fast bulk flows without introducing new parameters.

However, the agreement between the Newtonian and the quasi-Newtonian results is misleading because it simply manifests the fact that the latter analysis is also Newtonian (for all practical purposes). This happens because both approaches bypass (for different reasons) the gravitational input of the peculiar flux. In Newtonian physics this is unavoidable, since only the density of the matter gravitates. In the quasi-Newtonian studies, however, the effect of the peculiar flux is not accounted for because of the entirely unnecessary zero linear vorticity and shear constraints and the subsequent introduction of an (also unnecessary) scalar potential for the linear 4-acceleration. All this blurs the physics and diverts the attention from the key role of the peculiar flux, so that its contribution to the relativistic gravitational field and subsequently to the linear evolution of peculiar velocities are inadvertently bypassed. As a result, the linear evolution is identical to that of the purely Newtonian study.

Accounting for the gravitational contribution of the peculiar flux changes the picture drastically. The flux input to the energy-momentum tensor feeds to the relativistic conservation laws and then emerges in the linear evolution formula of the peculiar-velocity field, which differs profoundly from its Newtonian/quasi-Newtonian counterpart. The solution reveals a considerably stronger growth ($v\propto t$) for linear peculiar velocities between recombination and the dark-energy epoch. Moreover, a well known mathematical theorem guarantees that the above is the minimum growth-rate of the peculiar-velocity field. No restrictions have been imposed and no ad hoc assumptions have been made. It is pure general relativity and it could provide a simple, as well as physically motivated, answer to the bulk-flow puzzle. Moreover, it is also conceivable that this could be achieved without affecting the basics of the $\Lambda$CDM model, but by simply relaxing its current (Newtonian based) limits on large-scale peculiar motions.\\

\textbf{Acknowledgements:} This work was supported by the Hellenic Foundation for Research and Innovation (H.F.R.I.), under the ``First Call for H.F.R.I. Research Projects to support Faculty members and Researchers and the procurement of high-cost research equipment Grant'' (Project Number: 789).

\end{document}